\RequirePackage{fix-cm}
\documentclass[
twocolumn,
epjc3
]{svjour3}  

\usepackage{newtxtext,newtxmath} 
\usepackage{dcolumn}
\RequirePackage{graphicx}
\RequirePackage{subfigure}
\RequirePackage{rotating}
\RequirePackage{pdflscape}
\RequirePackage{color}
\usepackage{mathtools}
\usepackage{multirow}
\usepackage{extarrows}
\usepackage{booktabs}
\usepackage{microtype}
\usepackage{bm}
\RequirePackage{latexsym}
\RequirePackage{xspace}
\RequirePackage[numbers,sort&compress]{natbib}
\RequirePackage[colorlinks,citecolor=blue,urlcolor=blue,linkcolor=blue]{hyperref}
\usepackage{amsmath,amsfonts,bm}
\usepackage{wasysym}
\usepackage{color}
\usepackage{lineno}
\usepackage[usenames,dvipsnames,svgnames,table]{xcolor}	
\let\oldequation\equation
\let\oldendequation\endequation

\renewenvironment{equation}
  {\linenomathNonumbers\oldequation}
  {\oldendequation\endlinenomath}

\let\oldalign\align
\let\oldendalign\endalign

\renewenvironment{align}
  {\linenomathNonumbers\oldalign}
  {\oldendalign\endlinenomath}

\usepackage{acronym}
\usepackage{xcolor}

\newcommand{\qbb}{$Q_{\beta\beta}$\xspace}
\newcommand{\onu}{$0\nu\beta\beta$\xspace}
\newcommand{\nbb}{$2\nu\beta\beta$\xspace}

\newcommand{\kgy}{kg~$\times$~y\xspace}

\newcommand{\se}{$^{82}$Se\xspace}
\newcommand{\mo}{$^{100}$Mo\xspace}

\newcommand{\cd}{$^{116}$Cd\xspace}

\newcommand{\lmo}{Li$_2$MoO$_4$\xspace}

\newcommand{\lmoenr}{Li$_{2}${}$^{100}$MoO$_4$\xspace}

\newcommand{\mspeconea}{$\mathcal{M}_{1\alpha}$\xspace}
\newcommand{\mspeconeb}{$\mathcal{M}_{1\beta/\gamma}$\xspace}
\newcommand{\mspectwo}{$\mathcal{M}_2$\xspace}

\newcommand{\THO}{$^{232}$Th\xspace}
\newcommand{\URA}{$^{238}$U\xspace}
\newcommand{\CO}{$^{60}$Co\xspace}

\newcommand{\PO}{$^{210}$Po\xspace}

\newcommand{\SR}{$^{90}$Sr/$^{90}$Y\xspace}

\newcommand{\K}{$^{40}$K\xspace}

\newcommand{\onumajone}{$\beta\beta\chi_{0}$\xspace}
\newcommand{\onumajtwo}{$\beta\beta\chi_{0}\chi_{0}$\xspace}
\newcommand{\aof}{$\dot{a}_{o f}^{(3)}$\xspace}
\newcommand{\snbb}{$\nu N \beta\beta$\xspace}

\journalname{Eur. Phys. J. C}

\begin{document}

\title{Searching for Beyond the Standard Model physics using the improved description of $^{100}$Mo $2\nu\beta\beta$ decay spectral shape with CUPID-Mo}
\author{
C.~Augier\thanksref{IPNL}\and
A.~S.~Barabash\thanksref{ITEP}\and
F.~Bellini\thanksref{Sapienza,INFN-Roma}\and
G.~Benato\thanksref{GSSI,LNGS}\and
M.~Beretta\thanksref{UCB}\and
L.~Berg\'e\thanksref{IJCLab}\and
J.~Billard\thanksref{IPNL}\and
Yu.~A.~Borovlev\thanksref{NIIC}\and
L.~Cardani\thanksref{INFN-Roma}\and
N.~Casali\thanksref{INFN-Roma}\and
A.~Cazes\thanksref{IPNL}\and
E.~Celi\thanksref{e1,GSSI,LNGS}\and
M.~Chapellier\thanksref{IJCLab}\and
D.~Chiesa\thanksref{Milano,INFN-Milano}\and
I.~Dafinei\thanksref{INFN-Roma}\and
F.~A.~Danevich\thanksref{KINR, INFN-Roma2}\and
M.~De~Jesus\thanksref{IPNL}\and
T.~Dixon\thanksref{e2,IJCLab}\and
L.~Dumoulin\thanksref{IJCLab}\and
K.~Eitel\thanksref{KIT-IK}\and
F.~Ferri\thanksref{CEA-IRFU}\and
B.~K.~Fujikawa\thanksref{LBNLNSD}\and
J.~Gascon\thanksref{IPNL}\and
L.~Gironi\thanksref{Milano,INFN-Milano}\and
A.~Giuliani\thanksref{IJCLab}\and
V.~D.~Grigorieva\thanksref{NIIC}\and
M.~Gros\thanksref{CEA-IRFU}\and
D.~L.~Helis\thanksref{LNGS}\and
H.~Z.~Huang\thanksref{Fudan}\and
R.~Huang\thanksref{UCB}\and
L.~Imbert\thanksref{IJCLab}\and
A.~Juillard\thanksref{IPNL}\and
H.~Khalife\thanksref{CEA-IRFU}\and 
M.~Kleifges\thanksref{KIT-IPE}\and
V.~V.~Kobychev\thanksref{KINR}\and
Yu.~G.~Kolomensky\thanksref{UCB,LBNLNSD}\and
S.~I.~Konovalov\thanksref{ITEP}\and
J.~Kotila\thanksref{Jyv,Jyv2,yale}\and
P.~Loaiza\thanksref{IJCLab}\and
L.~Ma\thanksref{Fudan}\and
E.~P.~Makarov\thanksref{NIIC}\and
P.~de~Marcillac\thanksref{IJCLab}\and
R.~Mariam\thanksref{IJCLab}\and
L.~Marini\thanksref{LNGS}\and
S.~Marnieros\thanksref{IJCLab}\and
X.~F.~Navick\thanksref{CEA-IRFU}\and
C.~Nones\thanksref{CEA-IRFU}\and
E.~B.~Norman\thanksref{UCB}\and
E.~Olivieri\thanksref{IJCLab}\and
J.~L.~Ouellet\thanksref{MIT}\and
L.~Pagnanini\thanksref{GSSI,LNGS}\and
L.~Pattavina\thanksref{LNGS,TUM}\and
B.~Paul\thanksref{CEA-IRFU}\and
M.~Pavan\thanksref{Milano,INFN-Milano}\and
H.~Peng\thanksref{USTC}\and
G.~Pessina\thanksref{INFN-Milano}\and
S.~Pirro\thanksref{LNGS}\and
D.~V.~Poda\thanksref{IJCLab}\and
O.~G.~Polischuk\thanksref{KINR,INFN-Roma}\and
S.~Pozzi\thanksref{INFN-Milano}\and
E.~Previtali\thanksref{Milano,INFN-Milano}\and
Th.~Redon\thanksref{IJCLab}\and
A.~Rojas\thanksref{LSM}\and
S.~Rozov\thanksref{JINR}\and 
V.~Sanglard\thanksref{IPNL}\and
J.~A.~Scarpaci\thanksref{IJCLab}\and
B.~Schmidt\thanksref{CEA-IRFU}\and
Y.~Shen\thanksref{Fudan}\and
V.~N.~Shlegel\thanksref{NIIC}\and
F.~\v{S}imkovic \thanksref{CUB,CTUP}\and
V.~Singh\thanksref{UCB}\and
C.~Tomei\thanksref{INFN-Roma}\and
V.~I.~Tretyak\thanksref{KINR,LNGS}\and 
V.~I.~Umatov\thanksref{ITEP}\and
L.~Vagneron\thanksref{IPNL}\and
M.~Vel\'azquez\thanksref{UGA}\and
B.~Ware\thanksref{CU}\and
B.~Welliver\thanksref{UCB}\and
L.~Winslow\thanksref{MIT}\and
M.~Xue\thanksref{USTC}\and
E.~Yakushev\thanksref{JINR}\and
M.~Zarytskyy\thanksref{KINR}\and
A.~S.~Zolotarova\thanksref{CEA-IRFU}
}

\thankstext{e1}{Corresponding author: emanuela.celi@gssi.it}
\thankstext{e2}{Present address: Department of Physics \& Astronomy, University College London, Gower Street, London,
WC1E 6BT, UK}

\institute{Univ Lyon, Universit\'{e} Lyon 1, CNRS/IN2P3, IP2I-Lyon, F-69622, Villeurbanne, France  \label{IPNL} \and
National Research Centre “Kurchatov Institute”, Kurchatov Complex of Theoretical and Experimental
Physics, 117218 Moscow, Russia \label{ITEP} \and 
Dipartimento di Fisica, Sapienza Universit\`a di Roma, P.le Aldo Moro 2, I-00185, Rome, Italy \label{Sapienza} \and 
INFN, Sezione di Roma, P.le Aldo Moro 2, I-00185, Rome, Italy \label{INFN-Roma} \and
Gran Sasso Science Institute, L’Aquila I-67100, Italy \label{GSSI} \and
INFN, Laboratori Nazionali del Gran Sasso, I-67100 Assergi (AQ), Italy \label{LNGS} \and
Department of Physics, University of California, Berkeley, California 94720, USA \label{UCB} \and
Universit\'{e} Paris-Saclay, CNRS/IN2P3, IJCLab, 91405 Orsay, France \label{IJCLab} \and
Nikolaev Institute of Inorganic Chemistry, 630090 Novosibirsk, Russia \label{NIIC} \and
Dipartimento di Fisica, Universit\`{a} di Milano-Bicocca, I-20126 Milano, Italy \label{Milano} \and 
INFN, Sezione di Milano-Bicocca, I-20126 Milano, Italy \label{INFN-Milano} \and 
Institute for Nuclear Research of National Academy of Sciences of Ukraine, 03028 Kyiv, Ukraine \label{KINR} \and 
INFN, Sezione di Roma Tor Vergata, Via della Ricerca Scientifica 1, I-00133 Rome, Italy \label{INFN-Roma2} \and 
Institute for Astroparticle Physics, Karlsruhe Institute of Technology, 76021 Karlsruhe, Germany \label{KIT-IK} \and
IRFU, CEA, Universit\'{e} Paris-Saclay, F-91191 Gif-sur-Yvette, France  \label{CEA-IRFU} \and 
Nuclear Science Division, Lawrence Berkeley National Laboratory, Berkeley, California 94720, USA \label{LBNLNSD} \and
Key Laboratory of Nuclear Physics and Ion-beam Application (MOE), Fudan University, Shanghai 200433, PR China \label{Fudan} \and
Institute for Data Processing and Electronics, Karlsruhe Institute of Technology, 76021 Karlsruhe, Germany \label{KIT-IPE} \and
International Center for Advanced Training and Research in Physics (CIFRA), 409, Atomistilor Street, Bucharest-Magurele, 077125, Romania \label{Jyv} \and
Finnish Institute for Educational Research, University of Jyv\"askyl\"a, P.O. Box 35, FI-40014 Jyv\"askyl\"a, Finland \label{Jyv2} \and
Center for Theoretical Physics, Sloane Physics Laboratory, Yale University, New Haven, Connecticut 06520-8120, USA \label{yale}\and
Massachusetts Institute of Technology, Cambridge, MA 02139, USA \label{MIT} \and
Physik Department, Technische Universit\"at M\"unchen, Garching D-85748, Germany \label{TUM} \and
Department of Modern Physics, University of Science and Technology of China, Hefei 230027, PR China \label{USTC} \and
Universit\'{e} Grenoble Alpes, CNRS, Grenoble INP, LPSC/LSM-IN2P3, 73500 Modane, France \label{LSM} \and
Laboratory of Nuclear Problems, JINR, 141980 Dubna, Moscow region, Russia \label{JINR} \and
Faculty of Mathematics, Physics and Informatics, Comenius University in Bratislava, 842 48 Bratislava, Slovakia\label{CUB} \and
Institute of Experimental and Applied Physics, Czech Technical University in Prague, 128 00 Prague, Czech Republic\label{CTUP}\and
Universit\'e Grenoble Alpes, CNRS, Grenoble INP, SIMAP, 38402 Saint Martin d'H\'eres, France \label{UGA}\and
John de Laeter Centre for Isotope Research, GPO Box U 1987, Curtin University, Bentley, Western Australia, Australia\label{CU}
}




\date{Received: date / Accepted: date}

\maketitle

\begin{abstract}
The current experiments searching for neutrinoless double-$\beta$ (\onu) decay also collect large statistics of Standard Model allowed two-neutrino double-$\beta$ ($2\nu\beta\beta$) decay events. These can be used to search for Beyond Standard Model (BSM) physics via \nbb decay spectral distortions. \mo has a natural advantage due to its relatively short half-life, allowing higher \nbb decay statistics at equal exposures compared to the other isotopes. 
We demonstrate the potential of the dual read-out bolometric technique exploiting a \mo exposure of 1.47~\kgy, acquired in the CUPID-Mo experiment at the Modane underground laboratory (France). We set limits on \onu decays with the emission of one or more Majorons, on \nbb decay with Lorentz violation, and \nbb decay with a sterile neutrino emission. 
In this analysis, we investigate the systematic uncertainty induced by modeling the \nbb decay spectral shape parameterized through an improved model, an effect never considered before. This work motivates searches for BSM processes in the upcoming CUPID experiment, which will collect the largest amount of \nbb decay events among the next-generation experiments.
\end{abstract}

\section{Introduction}
\label{sec:intro}
\begin{figure*}
    \begin{center}
    \includegraphics[width=1.\textwidth]{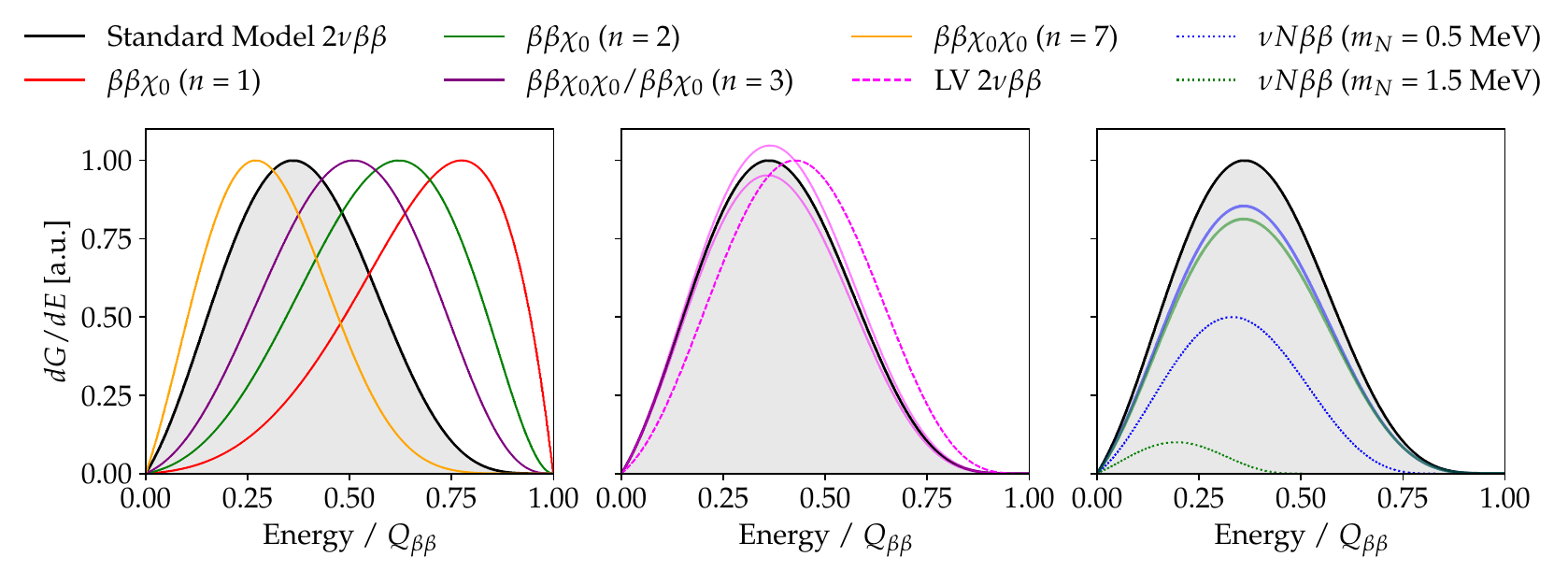}
    \caption{Standard Model \nbb decay (black) compared with different exotic double-$\beta$ decay spectra. Left panel: comparison of different Majoron-emitting modes with spectral indexes $n = 1,2,3$, and 7. Central panel: Lorentz-violating \nbb decay represented in the case of pure LV \nbb decay (dashed line) and summed to the SM \nbb one with an arbitrary value of $|$\aof$|$ (pink solid line). Right panel: \nbb decay with sterile neutrino emission (\snbb) for $m_N$ = 0.5~MeV (blue) and 1.5~MeV (green) in case of pure \snbb decay (dotted lines) and mixed with the SM \nbb one assuming $\sin^2 \theta = 0.1$ (blue and green solid lines).} 
    \label{fig:sp}
    \end{center}
\end{figure*}

Neutrinoless double-$\beta$ (\onu) decay is a hypothetical nuclear decay not allowed by the Standard Model (SM). The potential observation of \onu decay would demonstrate the violation of the $B-L$ symmetry of the SM. Moreover, it would also prove the Majorana nature of neutrinos, offering crucial insights into the fundamental symmetries governing particle interactions.
Beyond the primary focus on \onu decay, the pursuit of large masses, extended data collection periods, and appropriate $\beta\beta$ isotopes in \onu experiments lead to a high collection of events from two-neutrino double-$\beta$ (\nbb) decay~\cite{EXO-200:2013xfn,  GERDA:2023wbr, CUORE:2020bok, KamLAND-Zen:2019imh, Barabash:2018yjq, CUPID:2023wyy, NEMO-3:2019gwo, CUPID-Mo:2023lru}. \nbb decay is a second-order weak process occurring in the same \onu decay sources. It is particularly interesting for its relevance in the search for Beyond Standard Model (BSM) processes. Indeed, many theories predict the existence of exotic double-$\beta$ decays, called this way because they are characterized by a continuum energy distribution of the two emitted electrons similar to the \nbb one (Fig.~\ref{fig:sp})~\cite{Bossio:2023wpj}. 
These include the emission of new exotic particles such as scalar bosons known as ``Majorons'' or massive sterile neutrinos. Additionally, there is the possibility of observing violations of fundamental symmetries like the Lorentz invariance. Recently, other BSM cases have been investigated, like the potential effect of right-handed leptonic currents in \nbb~\cite{Deppisch:2020mxv} and the neutrino self-interactions~\cite{Deppisch:2020sqh}.

Majorons are massless Nambu-Goldstone bosons resulting from the spontaneous $B-L$ symmetry breaking in the low-energy regime~\cite{Gelmini:1980re, Chikashige:1980qk} and could play a role in the \onu decay coupling to the Majorana neutrinos~\cite{Georgi:1981pg}. Despite many theories being disfavored by the accurate measurements of the width of the Z boson decay~\cite{ALEPH:2005ab}, currently, different models predict the emission of one or two~\cite{Bamert:1994hb} Majorons in the \onu decay, these are denoted as
\begin{equation}
    (A, Z) \rightarrow(A, Z+2)+2 e^{-}+\chi_{0} \quad (\text{\onumajone}),
\end{equation}
or
\begin{equation}
    (A, Z) \rightarrow(A, Z+2)+2 e^{-}+2\chi_{0} \quad (\text{\onumajtwo}).
\end{equation}
In these models, the Majoron carries a non-zero lepton number~\cite{Burgess:1993xh} or it is a component of a massive gauge boson~\cite{Carone:1993jv} or a ``bulk'' field~\cite{Mohapatra:2000px}. Recently, schemes with massive Majoron-like particles have become popular since they could play the role of dark matter~\cite{Berezinsky:1993fm, Heeck:2017kxw, Brune:2018sab, Cepedello:2018zvr}.
The signature of these decays can be distinguished from the SM \nbb one thanks to the different spectral index $n$, which determines the position of the maximum intensity in the two emitted electrons' energy spectrum. In particular, the differential decay rate is proportional to
\begin{equation}
    \frac{d \Gamma}{d T} \propto \left(Q_{\beta \beta}-T\right)^{n},
\end{equation}
where $T$ is the total kinetic energy of the two electrons emitted, and $Q_{\beta \beta}$ is the Q-value. 
The value of the spectral index is $n = 5$ for the SM \nbb decay, while for the Majoron-emitting modes, it can be $n= 1,2,3$, or 7 (left panel Fig.~\ref{fig:sp}).

Some BSM theories involve a Lorentz invariance Violation (LV) and the CPT (Charge-Parity-Time reversal) symmetry violation terms in the Lagrangian. These theories have been developed in the Standard Model Extension (SME) framework, in such a way that the SM gauge invariance is preserved~\cite{Colladay:1996iz, Colladay:1998fq}. 
The neutrino sector was extensively studied in the SME framework~\cite{Kostelecky:2003xn, Kostelecky:2003cr, Kostelecky:2011gq} and the majority of these effects are experimentally investigated through neutrino oscillations and time-of-flight measurements~\cite{Kostelecky:2008ts}. 
Nevertheless, four operators equally change all neutrino energies and have no impact on oscillations, which can be studied through weak decays~\cite{Kostelecky:2008in, Diaz:2013saa} and are the so-called \textit{countershaded} operators, labeled as ``oscillation free'' (\textit{of}). 
The interaction of these operators with neutrinos modifies their four-momentum in a way that
\begin{equation}
    q^{\alpha}=(\omega, \mathbf{q}) \longrightarrow q^{\alpha}=\left(\omega, \mathbf{q}+\mathbf{a}_{o f}^{(3)}-\dot{a}_{o f}^{(3)} \hat{\mathbf{q}}\right),
\end{equation}
where $\mathbf{a}_{o f}^{(3)}$ encodes all the 3 directional \textit{of} components while \aof represents the anisotropic one~\cite{Kostelecky:2008in, Diaz:2013saa}. 
This results in a perturbation of the \nbb decay rate, which can be written as the sum of two components
\begin{equation}
    \Gamma=\Gamma_{SM}+ 10 \dot{a}_{o f}^{(3)} \Gamma_{LV},
\end{equation}
where $\Gamma_{SM}$ is the SM \nbb decay rate and $\Gamma_{LV}$ is the perturbation induced by the Lorentz violation. The energy distribution of the two electrons emitted in the LV \nbb decay is shifted to higher energies compared to the SM \nbb decay (central panel Fig.~\ref{fig:sp}). 

In the past years, some BSM theories have hypothesized the existence of a sterile neutrino $N$ with a mass $m_N$ at accessible energies~\cite{Chattopadhyay:2017zvs, Lee:2013htl, Shaposhnikov:2006nn, Deppisch:2010fr, Kersten:2007vk, Pilaftsis:2004xx, Bolton:2019pcu}.
The sterile neutrino can also be considered a candidate for dark matter~\cite{Dasgupta:2021ies}. Experimentally, the parameter of interest is the active-sterile mixing strength $\sin^2 \theta$, which determines the mixing angle between the electron neutrino flavor and the sterile one. 
The existence of a sterile neutrino with a mass $m_N < Q_{\beta\beta}$ can induce an effect on the \nbb decay spectral shape~\cite{Bolton:2020ncv, Agostini:2020cpz}, implying the emission of one sterile neutrino in the decay
\begin{equation}
    (A, Z) \rightarrow(A, Z+2)+2 e^{-}+\bar{\nu}+N \quad (\nu N \beta\beta),
\end{equation}
while the emission of two sterile neutrinos is considered negligible.
The \onu experiments are sensitive to the sterile neutrino mass range from 0.1~MeV up to 3~MeV (depending on \qbb) due to kinematical conditions. This region is particularly interesting due to the relative weakness of the actual constraints coming from single $\beta$-decay experiments, $\sin \theta \sim$~10$^{-3}$--10$^{-2}$~\cite{Holzschuh:2000nj, Schreckenbach:1983cg, Deutsch:1990ut, Borexino:2013bot, Derbin2018}. The effect on the total decay rate is given by 
\begin{equation}
    \Gamma= \cos^4 \theta \Gamma_{SM}+2 \cos^2 \theta \sin^2 \theta \Gamma_{\nu N},
\end{equation}
where the first term $\Gamma_{SM}$ accounts for the SM \nbb decay and the second term $\Gamma_{\nu N}$ represents the \snbb decay~\cite{Bolton:2020ncv, Agostini:2020cpz}. 
The energy spectrum in the case of pure \snbb decay events is characterized by a shift in the Q-value determined by the mass $m_N$ of the sterile state and a smaller Phase Space Factor (PSF) compared to \nbb decay (right panel Fig.~\ref{fig:sp}).

In this framework, scintillating cryogenic calorimeters are one of the most promising technologies for \onu decay searches \cite{Augier:2022znx, CUPID:2022puj} and the study of \nbb decay spectral shape~\cite{CUPID-Mo:2023lru, CUPID:2023wyy}. These detectors offer outstanding capabilities in terms of energy resolution, radiopurity, background rejection, detection efficiency, and mass scalability~\cite{CUORE:2021mvw, Augier:2022znx, CUPID-Mo:2023vle, CUPID:2022puj}. CUPID-Mo exploits this technology to demonstrate the performances and the experimental feasibility of a mid-scale experiment with \lmo-based bolometers.
In this paper, we present the results of the search for exotic double-$\beta$ decays of \mo exploiting the full data taking of the CUPID-Mo experiment. This analysis relies on the background model which provides a detailed description of the background sources releasing energy in the CUPID-Mo detector~\cite{CUPID-Mo:2023vle, CUPID-Mo:2023lru}.
Differently from the previous analyses, CUPID-Mo is the first experiment to account for the systematic uncertainties of the \nbb decay spectral shape, parameterized through the improved \nbb decay description~\cite{Simkovic:2018rdz, Nitescu:2021pdq}, in the search for exotic double-$\beta$ decays. The promising results obtained in this analysis motivate the interest in investigating the CUPID potential in the search for new physics processes involving distortion of the \nbb decay spectral shape.
CUPID is a next-generation ton-scale experiment aiming to reach unprecedented sensitivities on \onu decay of \mo using \lmoenr cryogenic calorimeters~\cite{CUPID:2019imh}. After one year of data taking, the \mo exposure in CUPID will be about 150 times higher than the CUPID-Mo, providing impressive statistics on the SM \nbb decay.  
Finally, with this work we analyze the main limits in searching for these BSM processes with cryogenic calorimeters and propose possible solutions to overcome these problems in future searches with CUPID.


\section{Experimental setup}
\label{sec:exp}
CUPID-Mo is an experiment at Laboratoire Souterrain de Modane (LSM) in France aiming to search for \onu decay of \mo. It was installed in the EDELWEISS cryostat~\cite{Armengaud:2017hit}, optimized for low background measurements. The detector acquired data at a stable temperature of $\sim$20~mK from March 2019 to June 2020, for a total \mo exposure of 1.47~\kgy.
The experiment is made up of 20 cryogenic calorimeters consisting of \lmoenr crystals produced from molybdenum enriched in $^{100}$Mo to $\sim$97\% and read-out by Neutron Transmutation Doped (NTD) Ge thermistors. Each crystal has a cylindrical shape and an average weight of 210~g.
\lmo-based bolometers are among the best detectors in this field due to the excellent energy resolution (7.7 $\pm$ 0.4 keV FWHM at 3034 keV~\cite{Augier:2022znx, Poda:2017jnl}) and the intrinsic radio-purity~\cite{CUPID-Mo:2023vle}. Moreover, these scintillating crystals permit the identification of the $\alpha$ background from the $\beta/\gamma$ interactions~\cite{Augier:2022znx}.
To detect the scintillation light and allow particle identification, thin Ge wafers, which also operate as cryogenic calorimeters, are positioned between \lmoenr crystals along the tower.
These Light Detectors (LDs) are also read out with NTD Ge thermistors and coated with a $\sim$70~nm layer of SiO$_2$ to increase the light collection. In this setup, each \lmoenr crystal faces the top and the bottom with LDs, as an exception for the crystals on the top floor which face only one LD.
A cylindrical copper holder and polytetrafluoroethylene pieces constitute the supporting structure for each crystal and the adjacent LDs. In addition, a reflective foil (3M~Vikuiti{\texttrademark}) surrounds each \lmoenr crystal to increase the light collection.
The detector set-up consists of 20 bolometers arranged in five towers, four modules for each tower. These are suspended using stainless steel springs to reduce the vibrational noise. 
More details on the detector structure are in Ref.~\cite{Armengaud:2019loe}. The cryogenic set-up is composed of five copper screens corresponding to the different thermal stages (300~K, 100~K, 50~K, 1~K, and 10~mK). Two lead shields aim to screen the detector from the environmental $\gamma$ radioactivity, an internal 14~cm thick Roman lead shield installed at the 1~K stage and a 20~cm thick external lead shield~\cite{EDELWEISS:2013wrh}. In the same way, two 10~cm and 55~cm thick polyethylene shields (internal and external, respectively) are installed to shield against environmental neutrons~\cite{Armengaud:2019loe}. Finally, the entire setup is surrounded by plastic scintillators acting as muon veto.

\section{Experimental data}
\label{sec:data}
Experimental data are acquired as a continuum time stream and digitized with a sampling frequency of 500 Hz both for \lmoenr detectors and LDs. The data are divided into twelve \textit{datasets}, where each  \textit{dataset} corresponds to about 1 month of data taking and is sub-divided into a series of \textit{runs}. Each \textit{run} is characterized by a period of $\sim$24 hours of stable data taking. At the beginning and the end of each \textit{dataset}, specific calibration \textit{runs} are acquired by deploying a \THO/\URA source in the vicinity of the detector array. The characteristic $\gamma$-lines of \THO and \URA produce several peaks in \lmoenr crystals enabling the calibration of all the detectors. The light detectors cannot be calibrated with the \THO/\URA sources since $\gamma$-rays are not fully contained in the LD volume. For this reason, specific \textit{runs} are acquired by using an intense \CO source, able to stimulate the production of fluorescence \mo X-rays at $\sim$17~keV used to calibrate LDs. 
Without calibration sources, each detector has an average trigger rate of about 14 mHz~\cite{CUPID-Mo:2023vle}.

Experimental data are acquired and processed with C++ based software packages developed by previous bolometric experiments~\cite{DiDomizio:2018ldc, Armengaud:2019loe}. All the triggered events are acquired in a 3-seconds time window (1 second of pre-trigger).
Each \lmoenr detector is associated with its adjacent LD(s).
First, we apply a series of cuts aiming to remove periods of detector instabilities and not-optimized data takings.
We reconstruct the pulse amplitudes using an optimal filter~\cite{Gatti:1986cw}. This method allows the production of a new pulse with the maximum signal-to-noise ratio by selectively weighting the frequency components of the signal and suppressing those that are more affected by noise. Then, the filtered waveforms are corrected in order to remove the amplitude's dependence on the initial detector temperature. For more details on the data processing see Ref.~\cite{Augier:2022znx}. The detectors' energy scale is determined through calibration \textit{runs}, which provide the calibration function parameters calculated using the $\gamma$-ray peaks produced by the \THO/\URA sources. A quadratic function with a zero intercept is used to calibrate the $\beta/\gamma$ energy region.
The $\alpha$ events appear in the background spectrum in a high energy range that spans from 4 to 10 MeV. Due to the different detector responses to $\gamma$-rays and $\alpha$-particles, the $\alpha$ spectrum shows a mis-calibration of about 8\%. In this case, we use the \PO $\alpha$-peak to re-calibrate the $\alpha$ region. Given the granularity of the detector, the information about the timing of each event with respect to the other detectors is extremely useful for studying the topology of radioactive decays, as explained in the next section. In order to do that, we tag each event occurring in a $\pm$10 ms time window with other detectors. The variable describing the number of channels that triggered a pulse in the same time window is called \textit{multiplicity}.

\section{Data selection}
The data selection for the source spectra employed in the background model fit (explained in Sec.~\ref{sec:BM}) is based on particle identification and time-coincidence criteria. 
The $\alpha$ identification relies on the scintillation light detected by the adjacent LD(s) coupled to each \lmoenr crystal~\cite{Augier:2022znx}.

Following the approach used in other bolometric experiments~\cite{CUORE:2016ons,CUPID:2023wyy}, we select the following spectra:
\begin{itemize}
    \item \mspeconea: $\alpha$ events with \textit{multiplicity = 1} and an energy in the 3 to 10~MeV range.
    \item \mspeconeb: $\beta/\gamma$ events with \textit{multiplicity = 1} and an energy in the 0.1 to 4~MeV range.
    \item \mspectwo: the sum of energies released in two crystals coincident in time (\textit{multiplicity = 2}) within the 0.2--4~MeV range. 
\end{itemize}
To select a clean data sample, we apply a series of quality cuts based on pulse shape and light yield. 
Details on the data selection and efficiency evaluation can be found in Refs.~\cite{CUPID-Mo:2023vle, Augier:2022znx}.
The resulting efficiencies from the cuts applied in CUPID-Mo data are $\varepsilon_1 = (88.9 \pm 1.1)$\% for \mspeconeb, $\varepsilon_2 = (83.3 \pm 2.5)$\% for \mspectwo, and $\varepsilon_3 = (94.7 \pm 1.0)$\% for \mspeconea.

\section{Monte Carlo simulations}
We reproduce the signature of the background sources by producing a series of Monte Carlo simulations. For this purpose, we employ the version 10.04 of \textsc{Geant4} \cite{Geant4}. The geometry implemented in simulations faithfully reproduces the experimental structure from small detector components to cryostat vessels and radiation shields (see Ref.~\cite{CUPID-Mo:2023vle} for more details). 
We generate radioactive decays in the experiment components using both Decay0~\cite{decay0} and \textsc{Geant4} (G4RadioactiveDecay library). Particle propagation through the experimental geometry employs the Livermore low-energy physics models~\cite{livermore}. 

We use the detector response model to reproduce the simulated spectra with the measured data. We model the energy resolution with a Gaussian shape where the mean is the energy deposited in the simulation and the standard deviation is derived from the experimental data. Pulses falling below the energy threshold of $<$40~keV are excluded, and the \textit{multiplicity} is reproduced within a $\pm$10~ms time window. We parameterize the scintillation light energy measured by the LD as a second-order polynomial of \lmoenr detectors energy to reconstruct light signals in data. The LD energy resolution is modeled as a Gaussian with a standard deviation depending on the energy deposited in the corresponding \lmoenr crystal. This approach ensures faithful replication of the light yield cuts in the simulations as observed in the experimental data. Data selection efficiencies are also taken into account in the simulations by generating a random number $r$ for each event uniformly distributed between 0 and 1. 

\section{Background model}
\label{sec:BM}
In our previous work, we conducted a thorough background model analysis utilizing a Markov Chain Monte Carlo (MCMC) Bayesian fit sampling the joint posterior probability density function (p.d.f.) of the model parameters utilizing the JAGS software~\cite{plummer2003jags, CUPID-Mo:2023vle}.
Experimental data for each bin $b$ and energy $E_b$ are modeled as a linear combination $f(E_b ; \vec{N})$ of the background sources weighted by a set of scaling factors called normalization parameters $\vec{N}$.
We consider as likelihood function the product of Poisson distributions
\begin{equation}
    \mathcal{L} (\operatorname{data}|\vec{N}) = \prod_{i}^{3} \prod_{b}^{N_{\text{bins}}} \operatorname{Pois}(n_{i,b} \mid f_{i} (E_b ; \vec{N}))
\end{equation}
where $n_{i,b}$ is the experimental number of counts for each bin in the \textit{i}-th spectrum. The joint posterior p.d.f. has the form
\begin{equation}
    p(\vec{N}|\operatorname{data}) \propto \mathcal{L} (\operatorname{data}|\vec{N}) \times \pi(\vec{N}),
\end{equation}
where $\pi(\vec{N})$ represents a set of the prior distributions~\cite{CUPID-Mo:2023vle}.
With a Bayesian fit, we can model the uncertainties of various background sources and include them directly in the fit by choosing a specific prior p.d.f. for each contribution.

In the list of background components utilized in the fit, we included \THO and \URA contaminations in all detector and cryostat components. 
We account for the breaks in the secular equilibrium of \THO and \URA sources by producing separate simulations of their sub-chains. Specifically, in the \THO chain, break points occur at $^{228}$Ra and $^{228}$Th, while in the \URA chain, break points are $^{234}$U, $^{230}$Th, $^{226}$Ra, and $^{210}$Pb. Additionally, we consider other contributions, such as \K contamination in the springs and the outermost cryogenic thermal shield, $^{60}$Co from cosmogenic activation in all copper components, and $^{40}$K, $^{87}$Rb and \SR in crystals. 
Decays are generated within the bulk of components and on the surface for nearby elements, following an exponential density profile $e^{-x/\lambda}$, where $\lambda$ is a variable depth parameter (for most of the surface contaminants $\lambda$~=~10~nm, see Ref.~\cite{CUPID-Mo:2023vle} for details). Initially, the \nbb decay of \mo was simulated in the background model under the Single-State Dominance (SSD) hypothesis, utilizing exact Dirac electron wave functions~\cite{Kotila:2012zza}. Subsequently, we incorporated an improved description~\cite{Simkovic:2018rdz, Nitescu:2021pdq} of the \nbb decay into the fit, with marginalization over the theoretical uncertainty of the spectral shape~\cite{CUPID-Mo:2023lru}. The fit also includes the \nbb decay of \mo to $^{100}$Ru 0$_1^+$ excited state, pile-up produced by \nbb decay events occurring in the same crystal, and random coincidences between two crystals.

Finally, a total of 67 sources are included in the fit. We modeled all the priors as non-negative uniform p.d.fs. The only informative priors are set for the \nbb decay from $^{100}$Mo to $^{100}$Ru 0$_1^+$ excited state (with a half-life of $T_{1/2}=(6.7 \pm 0.5) \times 10^{20}$~y~\cite{Barabash:2b2n}), the stainless steel springs contamination (from screening measurements), the accidental coincidences, determined from the rate of single events, and the \nbb decay pile-up, estimated with the calibration \textit{runs} (see Ref.~\cite{CUPID-Mo:2023vle} for details). Variable binning is employed for the three spectra to ensure sufficient counts in each bin, thereby minimizing the impact of statistical fluctuations. A minimum bin size of 15~keV is chosen for \mspeconeb and \mspectwo, while a 20~keV minimum bin size is used for \mspeconea. The minimum number of counts in each bin is 50 for \mspeconeb and 30 for \mspectwo and \mspeconea. Additionally, each peak is selected to be fully contained within one bin to mitigate the systematic effects of the detector response on the results. 

\section{BSM analysis}
\label{sec:analysis}
The search for exotic double-$\beta$ decays is performed by individually incorporating the new physics spectra into the background model fit.
Each MCMC sample reconstructs the \textit{joint posterior} p.d.f. of the normalization parameters. The normalization parameter $N_j$ of the j-th BSM process is directly related to the decay rate through the formula
\begin{equation}
    \Gamma_j = \dfrac{N^{j}_{MC} \cdot N_{j}}{T \cdot \varepsilon \cdot N_{\beta\beta}}.
\end{equation}
Here, $N^{j}_{MC}$ denotes the total number of decays generated in the simulation, $N_{\beta\beta}$ is the total number of \mo atoms, $T$ is the experiment lifetime and $\varepsilon$ is the efficiency. In our case, the efficiency has been included in the simulations and, therefore, already accounted into the normalization parameter. The value of the product $T \cdot N_{\beta\beta}$ in CUPID-Mo corresponds to $(99.7 \pm 0.2)\times 10^{23}$ \mo nuclei~$\times$~y.


All new physics spectra in this analysis are simulated using exact Dirac wave functions with finite nuclear size and electron screening~\cite{JKprivate}. We use the SSD approximation for all the BSM spectra since the difference with their corresponding higher-state dominance (HSD) approximated spectrum in the total number of events is negligible compared to the systematic uncertainties of the model. 
We define as \textit{reference} fit the configuration of sources described above. We consider two possible scenarios: the ``standard'' background model fit, as described in Ref.~\cite{CUPID-Mo:2023vle}, which uses the SSD hypothesis for \mo \nbb decay, and the Improved Model (IM) fit, detailed in Ref.~\cite{CUPID-Mo:2023lru}, which utilizes the improved description for the \mo \nbb decay spectral shape~\cite{Simkovic:2018rdz, Nitescu:2021pdq}. Previous analyses from many experiments in this field use either the SSD or the HSD approximation to model the \nbb decay spectral shape~\cite{GERDA:2022ffe, CUPID-0:2022yws, NEMO-3:2015jgm,NEMO-3:2019gwo, Barabash:2018yjq, Kharusi:2021jez,KamLAND-Zen:2012uen,Armengaud:2019rll}. 
For the first time in the search for exotic double-$\beta$ decays, the uncertainties of the SM \nbb decay spectral shape are marginalized in the fit using an extended model to describe the \nbb spectral shape~\cite{Simkovic:2018rdz, Nitescu:2021pdq}.
Following the improved description, the \nbb decay rate or, more specifically, the PSF can be decomposed as
\begin{equation}
        G^{2 \nu} = G_{0}^{2 \nu} + \xi_{31}^{2 \nu} G_{2}^{2 \nu} + \frac{1}{3}\left(\xi_{31}^{2 \nu}\right)^{2} G_{22}^{2 \nu} + \left[\frac{1}{3}\left(\xi_{31}^{2 \nu}\right)^{2}
                           + \xi_{51}^{2 \nu}\right] 
                           G_{4}^{2 \nu},
\end{equation}
where $G_{0}, G_{2}, G_{22}$ and $G_{4}$ are the PSFs for different terms in the Taylor expansion of the lepton energies and $\xi_{31}$, $\xi_{51}$ are parameters depending on the ratios of the Gamow-Teller Nuclear Matrix Elements (NMEs). For \mo, the values of the PSFs are $G_{0}$~=~3.303~$\times$~$ 10^{-18}$~y$^{-1}$, $G_{2}$~=~1.509~$\times$~$10^{-18}$~y$^{-1}$, $G_{22}$~=~4.320~$\times$~$10^{-19}$~y$^{-1}$, and $G_{4}$~=~1.986~$\times$~$10^{-19}$~y$^{-1}$~\cite{Simkovic:2018rdz}. 
In the IM fit, the four spectra corresponding to $G_{0}, G_{2}, G_{22}$, and $G_{4}$ are simulated using exact Dirac wave functions with finite nuclear size and electron screening~\cite{JKprivate}, and they are included in the fit separately. To accurately describe the \nbb decay spectral shape, the fit model is modified to marginalize over the $\xi$ parameters. Since $\xi_{31}$ and $\xi_{51}$ are strongly anti-correlated, a Gaussian prior is placed on the ratio $\xi_{31}/\xi_{51}$ with a mean equal to the SSD prediction and a conservative 5\% uncertainty~\cite{CUPID-Mo:2023lru}. This choice relies on the nuclear structure calculations, where the value of $\xi_{31}/\xi_{51}$ can be reliably obtained~\cite{Simkovic:2018rdz, Coraggio:2022vgy}. Within the SSD hypothesis, the value of the ratio is 0.367~\cite{Simkovic:2018rdz}. For more details on the IM fit, see Ref.~\cite{CUPID-Mo:2023lru}.

We perform a series of additional fits to assess the systematic uncertainties. These are listed as follows:
\begin{itemize}
    \item The dominant contribution at low energies comes from \THO and \URA contaminations in cryostat components. However, the signatures of these sources in close (10~mK) and far (300~K) components are almost degenerate, leading to a possible mis-modeling. To estimate the uncertainty related to the \textbf{source location}, we alternatively remove the 300~K and 10~mK sources from the fit.
    \item In the background model fit, the decay chain \textbf{\SR} is included. Both these isotopes decay through pure $\beta$-decays, producing a featureless spectrum correlating with the \nbb decay and BSM components. In the background model fit we measure a \SR activity of $179^{+36}_{-32}$~$\mu$Bq/kg. However, the presence of \SR is still uncertain since other unexplained contributions at low energies can induce its convergence. This systematic test involves removing this contribution from the fit.
    \item In the \textit{reference} fit, some contributions present a posterior p.d.f. converging to a value compatible with zero or show an exponential shape flattened to zero. The \textbf{minimal model} is a fit performed by removing these contributions.
    \item As described above, we use a minimum bin size of 15~keV on \mspeconeb, nevertheless, we consider also different values for the \textbf{binning} (1~keV, 2~keV, and 20~keV).
    \item Two fits are performed varying the energy scale by $\pm$1~keV to account for a possible \textbf{energy bias}.
    \item The theoretical uncertainties on the \nbb decay \textbf{Bremss\-trahlung} cross-section may affect the accuracy of MC simulations on the \nbb decay spectrum. In order to assess this uncertainty, we perform the fit with alternative \nbb decay spectra obtained by varying the Bremsstrahlung cross-section by $\pm$10\%~\cite{Pandola:2014uea}.
\end{itemize}
The uncertainties on the efficiency (1.2\%) and the \mo enrichment (0.2\%) are directly marginalized in the posterior p.d.fs. with Gaussian priors. 
In the following sections, the way to extract the physical parameters for the different BSM processes and their results are described. 

\section{Results}
\label{sec:res}
Even though the search for all the exotic double-$\beta$ decays is conducted with the same fitting procedure, the way to extract a limit on the physical parameters is significantly different. Majoron-emitting decays do not interfere with the SM \nbb decay which is just considered an independent process. In this case, the SM \nbb decay acts as a background. 
Conversely, the existence of \snbb decay tends to suppress the SM \nbb decay rate by a factor $\cos^4 \theta$. 
The Lorentz-violating \nbb decay also interferes with the SM \nbb decay rate by introducing an additional term on the PSF. 
In the last two scenarios, increasing the sample size of \nbb decay events will result in lower statistical uncertainties. This will make it more sensitive to deviations~\cite{Bossio:2023wpj}. \mo is one of the best nuclides among the \nbb decay emitters to detect Lorentz-violating \nbb decay and sterile neutrino emissions because of its relatively short half-life.
\subsection{Majoron-emitting decays}
\begin{figure*}
    \begin{center}
    \includegraphics[width=.7\textwidth]{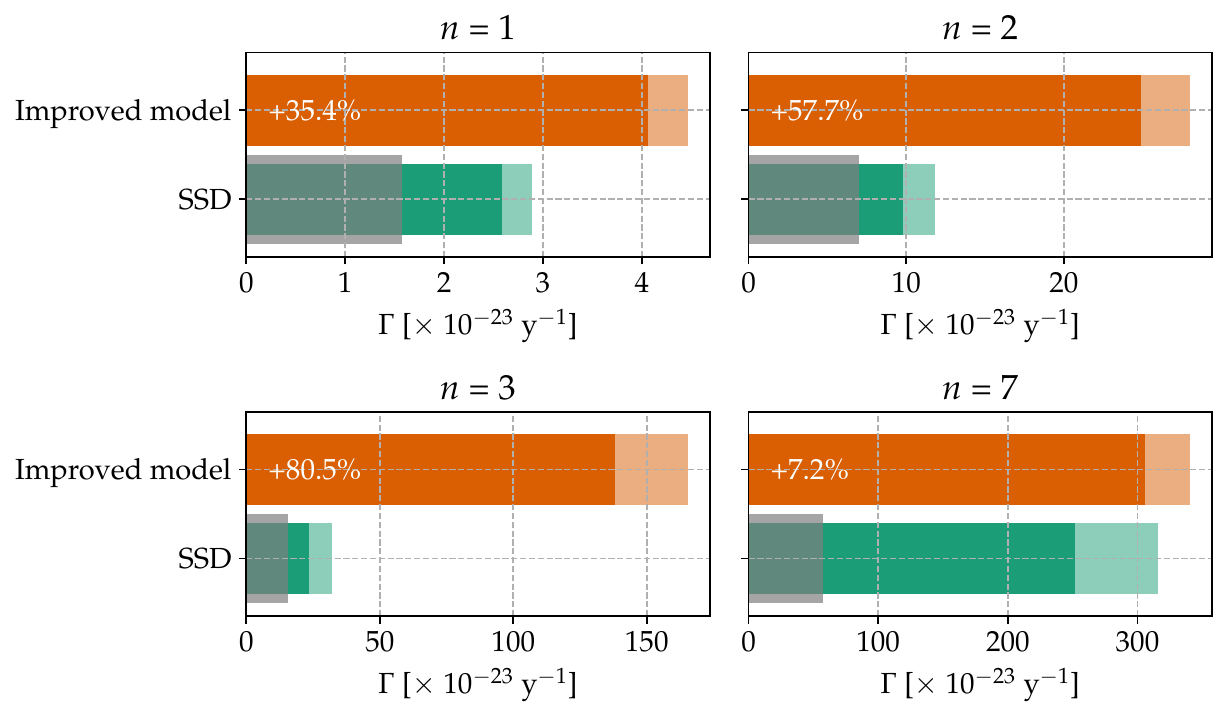}
    \caption{Upper limits at 90\% CI on the decay rate of different Majoron-emitting decays with the SSD approximation for \nbb decay (green) and the improved description implemented in the fit (orange). The solid color corresponds to the limit obtained from the \textit{reference} while the shaded one indicates the systematic uncertainty. The percentage difference between the results using the SSD and the IM fits (with the systematics) is highlighted in the orange column. 
    The shaded gray columns show the limit obtained by NEMO-3 using the SSD assumption~\cite{NEMO-3:2015jgm,NEMO-3:2019gwo}.} 
    \label{fig:majim}
    \end{center}
\end{figure*}

In this study, we performed the analysis of four different Majoron-emitting modes, corresponding to $n = 1,2,3$, and 7. The parameter of interest is the decay rate $\Gamma_{0 \nu M}$ of the different processes, which can be easily converted into a limit on the neutrino-Majoron coupling $g_{ee}^{M}$ through the formula 
\begin{equation}
    \Gamma_{0 \nu M}
    =
    G_{0 \nu M}
    \left| \left\langle g_{ee}^{M} \right\rangle \right| ^{2 m}
    \left|M_{0 \nu M}\right|^{2},
\end{equation}
where $G_{0 \nu M}$ and $\left|M_{0 \nu M}\right|$ are PSFs and NMEs for the Majoron-emitting modes, while $m$ is the number of Majorons emitted. The \mo values of $G_{0 \nu M}$ and $\left|M_{0 \nu M}\right|$ for different Majoron-emitting modes are summarized in Table~\ref{tab:majpsf}.
\begin{table}
    \centering
    \caption{\mo values of the PSF $G$ and NME for different Majoron-emitting decays. The values of $G$ are taken from Ref.~\cite{Kotila:2015ata}. For the Majoron mode with $n$ = 1, the NME values are taken from Ref.~\cite{Agostini:2022zub}. For the other modes, the NMEs have been calculated in the framework of the interacting boson model~\cite{Kotila:2021mtq}. For the Majoron mode $n$ = 2, no calculations for the PSF and the NME are available.}
    \label{tab:majpsf}
    \begin{tabular}{cccc}
        \toprule
        decay mode  & $n$             & $G$ [$\times 10^{-18}$~y$^{-1}$]          & NME  \\
        \midrule
        \onumajone  &1    & 598           & 3.84--6.59 \\
        \onumajone  &2    & --             & --               \\ 
        \onumajone  &3    & 2.42          & 0.263 \\
        \onumajtwo  &3    & 6.15          & 0.0019  \\    
        \onumajtwo  &7    & 50.8          & 0.0019 \\
        \bottomrule
    \end{tabular}
\end{table}
No evidence of signal was found for any of the decays mentioned above setting 90\% Credible Interval (CI) limits on their half-lives by integrating the posterior p.d.f. on the corresponding decay rates.
We performed the aforementioned systematic tests to determine the systematic uncertainty. For $n = 1, 2$, and 3, the systematics with a greater impact on the results are +10\% Bremsstrahlung, --1~keV energy scale, and \SR. In the Majoron mode $n = 7$, where the signature exhibits a spectrum shifted at lower energies compared to the SM \nbb decay (see Fig.~\ref{fig:sp}), the situation is inverted, with dominant effects from --10\% Bremsstrahlung, +1~keV energy scale, and source location. These tests demonstrate that uncertainties in modeling the low-energy part of the \mspeconeb spectrum limit the sensitivity for exotic double-$\beta$ decay searches.
In a conservative approach, we quote as the final limit the least stringent including systematics.
The same systematic checks are performed in the IM fit. The uncertainty on the \nbb decay spectral shape, parameterized in the improved description, significantly reduces the sensitivity for Majoron-emitting decays, as depicted in Fig.~\ref{fig:majim}. This effect is more pronounced in Majoron modes with $n < 5$, where the signal relies mostly on the high-energy side of the \nbb decay spectrum.
We report the limits on half-lives and neutrino-Majoron coupling constants obtained in this analysis for SSD and IM in Table~\ref{tab:majlim}.
Only results with the SSD fit can be directly compared with the other experiments, where the same assumption on the \nbb decay spectral shape was made. In the SSD assumption, the obtained limits are less stringent with factors of 1.8 for $n$ = 1, 1.7 for $n$ = 2, 2 for $n$ = 3, and 5.4 for $n$ = 7 compared to NEMO-3~\cite{NEMO-3:2015jgm, NEMO-3:2019gwo}.
However, the \mo exposure available in CUPID-Mo ($\sim$1.5~\kgy) is 22 times less than the NEMO-3 exposure ($\sim$34~\kgy), demonstrating the high sensitivity of the dual readout bolometric technique for these searches. 
\begin{table*}
    \centering
    \caption{Lower limits on the half-life at 90\% CI and upper limits on the neutrino-Majoron coupling constant for different Majoron-emitting modes are shown both for SSD and IM fits, including systematics. 
    }
    \label{tab:majlim}
    \begin{tabular}{cccccc}
        \toprule
        \multirow{2}*{Decay mode}   & \multirow{2}*{$n$}        & \multicolumn{2}{c}{$T_{1/2}$}         & \multicolumn{2}{c}{$g_{ee}^M$}      \\
                                    &                            &  limit SSD [y]    & limit IM [y]      & limit SSD    & limit IM   \\
        \midrule
        \onumajone & 1	    &$>$ 2.4 $\times$ 10$^{22}$	 &$>$ 1.6 $\times$ 10$^{22}$ & $<$ (4.0--6.9) $\times$ $10^{-5}$ & $<$ (5.0--8.5) $\times$ $10^{-5}$ \\
        \onumajone & 2     &$>$ 5.8 $\times$ 10$^{21}$	 &$>$ 2.7 $\times$ 10$^{21}$ & -- & -- \\
        \onumajone & 3     &$>$ 2.2 $\times$ 10$^{21}$	 &$>$ 0.5 $\times$ 10$^{21}$ & $<$ 0.053 & $<$ 0.112 \\
        \onumajtwo & 3     &$>$ 2.2 $\times$ 10$^{21}$	 &$>$ 0.5 $\times$ 10$^{21}$ & $<$ 2.1 & $<$ 3.1 \\
        \onumajtwo & 7     &$>$ 2.2 $\times$ 10$^{20}$	 &$>$ 2.0 $\times$ 10$^{20}$ & $<$ 2.2 & $<$ 2.3  \\
        \bottomrule
    \end{tabular}
\end{table*}

\subsection{Lorentz-violating \nbb decay}
The violation of Lorentz and CPT symmetries introduces a perturbation in the SM-\nbb spectrum. 
By adding the LV spectrum in the background model, the measured decay rate of the LV perturbation $\Gamma_{LV}^m$ and the SM \nbb decay $\Gamma_{SM}^m$ can be evaluated from the normalization parameters. Since the \textit{countershaded operator} (\aof) can assume negative values, under-fluctuations for the LV component are allowed in the fit.
The ratio of the decay rates is directly proportional to \aof through
\begin{equation}
    \dfrac{\Gamma_{LV}^m}{\Gamma_{SM}^m} = \dot{a}_{o f}^{(3)} \cdot 10 \cdot \dfrac{\delta G_{LV}}{G_{SM}},
\end{equation}
where the NMEs cancel out, $\delta G_{LV}$ is the PSF of the Lorentz perturbation, and $G_{SM}$ is the SM \nbb decay PSF. Finally, the \textit{countershaded operator} can be calculated from
\begin{equation}
    \dot{a}_{o f}^{(3)} = C \cdot \dfrac{\Gamma_{LV}^m}{\Gamma_{SM}^m},
\end{equation}
where $C$ is a constant value. In this case, the collection of high \nbb decay statistics plays an important role in constraining possible Lorentz-violating effects.
In both the SSD and the IM fit, the posterior p.d.f. converges within a range compatible with zero, setting a double-sided limit at 90\% CI on the negative and positive values of the \aof parameter.
We set the limits by taking into account the strong anti-correlation of the LV \nbb decay component with the SM one by calculating the $\dfrac{\Gamma_{LV}^m}{\Gamma_{SM}^m}$ ratio for each MCMC sampling. In the SSD fit, the value of the factor $C$ is calculated from the values of the PSFs in the SSD assumption, $C = (299.0$~$\pm$~$0.2)$~$\times$~$10^{-6}$~GeV$^{-1}$~\cite{JKprivate}, where the error comes mostly from the uncertainty on the \mo \qbb and it is calculated as in Ref.~\cite{Nitescu:2020xlr}. In the IM fit, the PSF of SM \nbb decay is explicitly calculated from the $\xi$ parameters.
\begin{figure}
    \begin{center}
    \includegraphics[width=.45\textwidth]{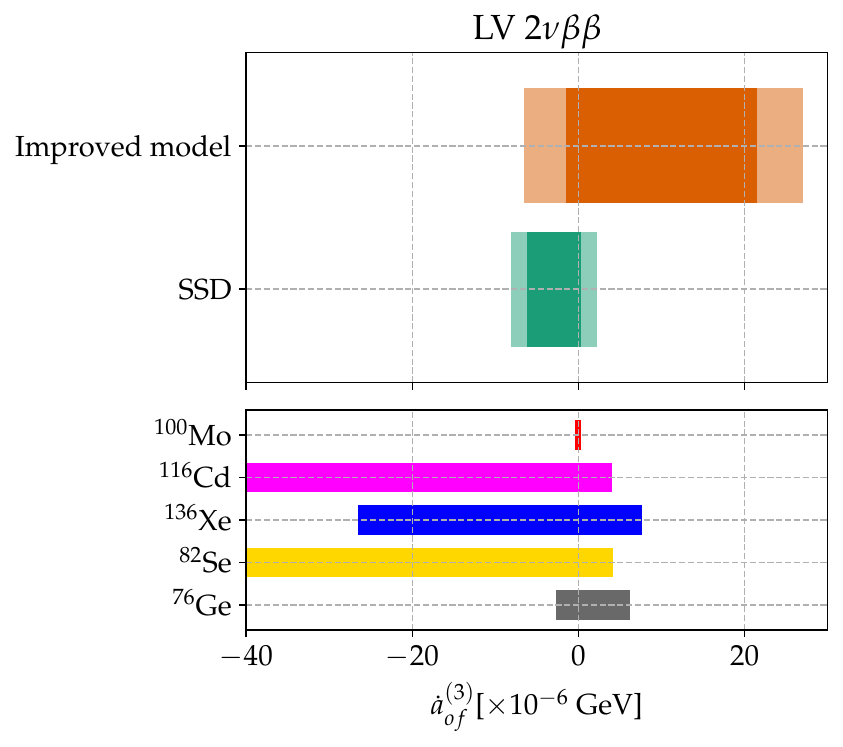}
    \caption{
    Top: double-sided limits at 90\% CI on the Lorentz-violating \textit{countershaded operator} with the SSD approximation for \nbb decay (green) and the improved description implemented in the fit (orange). The solid color corresponds to the limit obtained from the \textit{reference} fit, while the shaded one shows the effect of systematics. Bottom: current limits on the Lorentz-violating \textit{countershaded operator} for different double-$\beta$ decay emitters at 90\% CL obtained assuming either SSD or HSD approximations~\cite{GERDA:2022ffe,CUPID:2019kto,EXO-200:2016hbz,NEMO-3:2019gwo,Barabash:2018yjq}. For \se and \cd only positive limits are provided.}
    \label{fig:cptim}
    \end{center}
\end{figure}
The systematic checks listed above are performed on both the SSD and the IM fits, showing dominant effects for Bremsstrahlung, energy scale, and \SR. A summary of the results is represented in Fig.~\ref{fig:cptim}.
The final limits including systematic in the SSD fit are 
\begin{equation*}
    -8.1\times 10^{-6} < \dot{a}_{o f}^{(3)} < 2.2\times 10^{-6},
\end{equation*}
while in the IM fit
\begin{equation*}
    -6.5\times 10^{-6} < \dot{a}_{o f}^{(3)} < 2.5\times 10^{-5}.
\end{equation*}
The strong anti-correlation between the LV spectrum and the $\xi$ parameters leads to a large broadening of the \aof posterior, thus a weaker limit. The most stringent limit on the LV \nbb decay of \mo has been set by NEMO-3~\cite{NEMO-3:2019gwo}, corresponding to $(-4.2<\dot{a}_{o f}^{(3)}<3.5) \times 10^{-7}$ (Fig.~\ref{fig:cptim}), assuming the single state dominance for the \nbb decay. In our case, the SSD fit prefers negative values of the \aof parameter, producing a stronger limit on the positive values than the negative ones. Contrarily, in the IM fit the \aof converges at positive values. 

\subsection{Sterile neutrino emissions}
\begin{figure}
    \begin{center}
    \includegraphics[width=.45\textwidth]{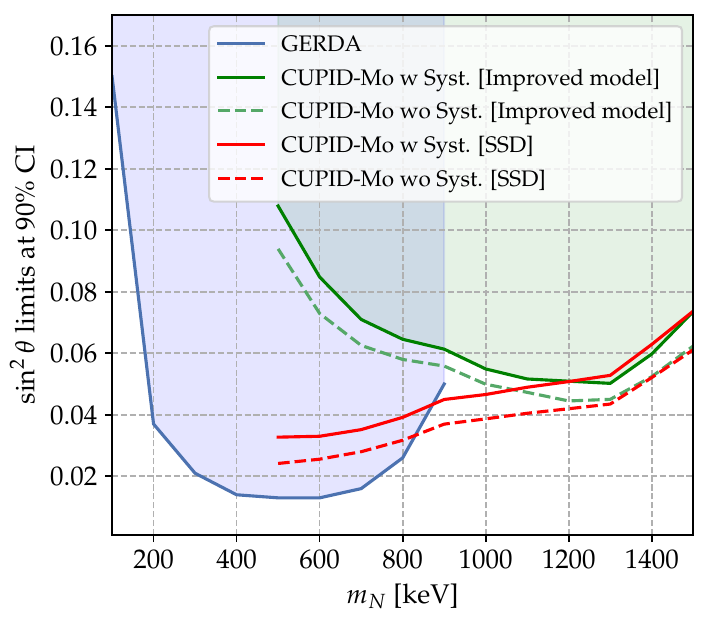}
    \caption{Limits on the active-sterile mixing strength $\sin^2 \theta$ as a function of the sterile neutrino mass $m_N$. The blue region represents the region excluded by the GERDA including systematics, covering the $m_N$ range 0.1--0.9~MeV~\cite{GERDA:2022ffe}. The green region is excluded by CUPID-Mo assuming the improved model to describe the \nbb decay spectral shape and including systematics. The green dashed line shows the limit obtained from the reference IM fit. The red lines represent the limits obtained using the SSD assumption with and without systematics (solid and dashed lines, respectively).} 
    \label{fig:SNmass}
    \end{center}
\end{figure}

We investigated a mass range for sterile neutrino masses $m_N$ from 0.5~MeV to 1.5~MeV, with 0.1~MeV steps. The active-sterile mixing strength can be calculated for different sterile neutrino masses as~\cite{Bolton:2020ncv, Agostini:2020cpz}
\begin{equation}
    \sin^2 \theta= \frac{G_{SM}}{2 G_{\nu N}} \cdot \dfrac{\Gamma_{\nu N}^m}{\Gamma_{SM}^m},
\end{equation}
where $G_{\nu N}$ is the PSF of the sterile neutrino spectrum, calculated for each sterile neutrino mass considered~\cite{JKprivate}. The same considerations on $G_{SM}$ done in the previous section are also valid in this case. No signal evidence is found for any of the considered masses, setting a 90\% CI limit on $\sin^2 \theta$. We identified as dominant systematic effects --10\% Bremsstrahlung, +1~keV energy scale, and minimal model. The same systematics are performed in the IM fit. All the obtained limits are summarized in Table~\ref{tab:snlim} and Fig.~\ref{fig:SNmass}. 
For sterile neutrino masses lower than 1.2 MeV, the \snbb decay component starts correlating with the \nbb one, and this effect is amplified when the \nbb decay shape is described with the IM. In particular, for low values of $m_N$ the shape of \snbb decay spectrum becomes more similar to the SM component. In the framework of \onu experiments, GERDA set limits on $\sin^2 \theta$, spanning a range for sterile neutrino masses from 0.1 to 0.9 MeV, as reported in Fig.~\ref{fig:SNmass}.
Nevertheless, existing bounds on the active-sterile mixing strength from $\beta$-decay and solar neutrino experiments~\cite{Holzschuh:2000nj, Schreckenbach:1983cg, Deutsch:1990ut, Borexino:2013bot, Derbin2018} have already excluded this region of the parameter space, setting limits on $\sin \theta$ in the range 10$^{-3}$--10$^{-2}$~\cite{Holzschuh:2000nj, Schreckenbach:1983cg, Deutsch:1990ut, Borexino:2013bot, Derbin2018}.
\begin{table}
    \centering 
    \caption{Upper limits on the active-sterile mixing strength $\sin^2 \theta$ for different values of sterile neutrino masses $m_N$~\cite{JKprivate}.}
    \label{tab:snlim}
    \begin{tabular}{ccccc}
        \toprule
        $m_N$ [MeV] & PSF [$\times 10^{-18}$ y$^{-1}$] & Limit SSD & Limit IM & diff.\\
        \midrule
        0.5	 & 2.19	 & $<$ 0.033	 & $<$ 0.108	 & +229\% \\
        0.6	 & 1.87	 & $<$ 0.033	 & $<$ 0.085	 & +156\% \\
        0.7	 & 1.57	 & $<$ 0.035	 & $<$ 0.071	 & +101\% \\
        0.8	 & 1.28	 & $<$ 0.039	 & $<$ 0.065	 & +64\% \\
        0.9	 & 1.03	 & $<$ 0.045	 & $<$ 0.061	 & +36\% \\
        1.0	 & 0.91	 & $<$ 0.047	 & $<$ 0.055	 & +17\% \\
        1.1	 & 0.81	 & $<$ 0.049	 & $<$ 0.052	 & +5.5\% \\
        1.2	 & 0.71	 & $<$ 0.051	 & $<$ 0.051	 & +0.2\% \\
        1.3	 & 0.62	 & $<$ 0.053	 & $<$ 0.050	 & $-$4.9\% \\
        1.4	 & 0.47	 & $<$ 0.063	 & $<$ 0.060	 & $-$5.0\% \\
        1.5	 & 0.34	 & $<$ 0.074	 & $<$ 0.074	 & $-$0.2\% \\
        \bottomrule
    \end{tabular}
\end{table}
\begin{figure*}[ht!]
    \begin{center}
    \includegraphics[width=.8\textwidth]{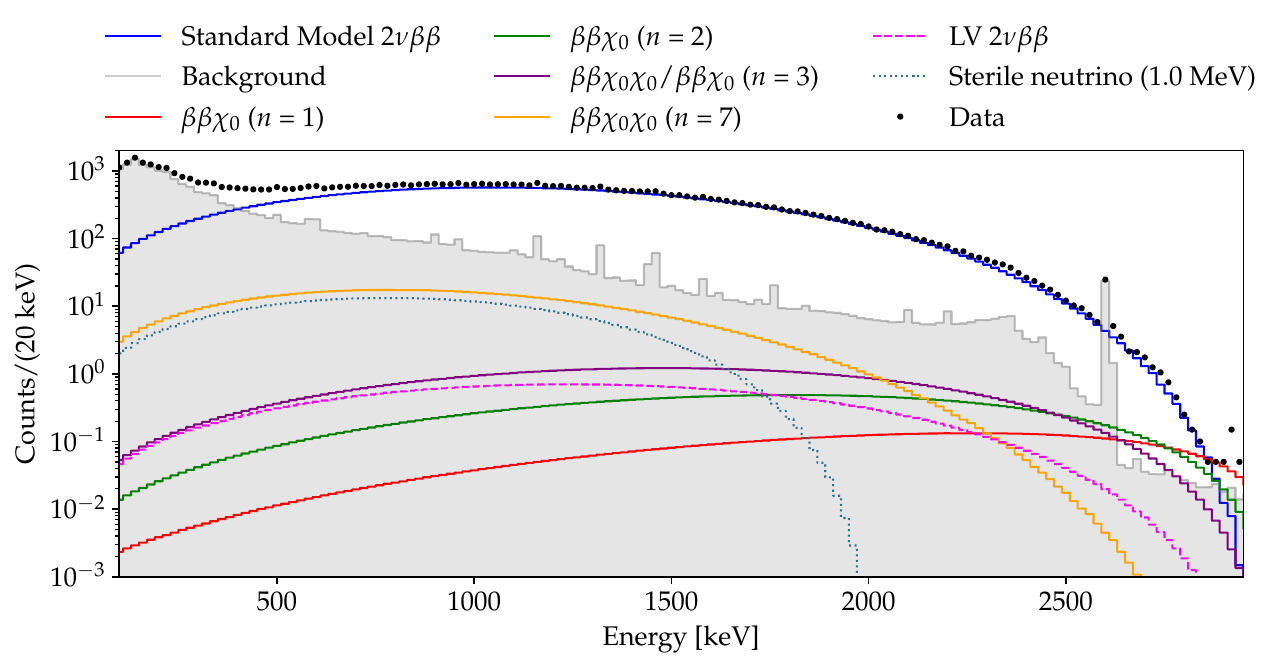}
    \caption{Exotic double-$\beta$ decay spectra compared to the experimental data with a number of counts corresponding to the 90\% CI limit obtained in the SSD fit. The LV \nbb decay spectrum represents only the limit on the positive side. The black dots correspond to the experimental data, the gray spectrum is the background reconstructed by the background model while the blue spectrum represents the SM \nbb decay.} 
    \label{fig:limitsData}
    \end{center}
\end{figure*}

\section{Conclusions and outlook}
We presented the results of the searches for exotic double-$\beta$ decays with CUPID-Mo. The analysis exploited the precise spectral shape reconstruction provided by the background model. 
No signal evidence has been found for any of the BSM processes investigated, setting a 90\% CI limit on the corresponding new physics parameter. For the first time, the theoretical uncertainties of the \nbb decay spectral shape, parameterized through the improved model description~\cite{Simkovic:2018rdz}, have been taken into account in this type of search.
This work demonstrated that uncertainties on the \nbb decay shape induce a significant reduction in sensitivity for all the processes investigated, requiring better theoretical constraints, higher statistics, and precise background reconstruction at low energies in the next-generation experiments. A parallel analysis was performed using a fixed spectral shape for the \nbb decay (single-state dominance) to compare the results with those of other experiments.
The limits at 90\% CI on the experimental data are represented in Fig.~\ref{fig:limitsData}. CUPID-Mo set stringent constraints on the neutrino-Majoron coupling and the Lorenz-violating \textit{countershaded operator}, despite the relatively small exposure ($\sim$1.5~\kgy), the limits are only a factor 2--10 less stringent than NEMO-3 ones ($\sim$34~\kgy)~\cite{NEMO-3:2015jgm, NEMO-3:2019gwo}. The search for \snbb decay has been performed for the first time using cryogenic calorimeters.
Exploiting the high Q-value of \mo, CUPID-Mo data allowed constraining the active-sterile mixing strength for higher values of $m_N$ compared to GERDA. Nevertheless, that region of the parameter space was already excluded by $\beta$-decay and solar neutrino experiments~\cite{Holzschuh:2000nj, Schreckenbach:1983cg, Deutsch:1990ut, Borexino:2013bot, Derbin2018}.
The results of CUPID-Mo demonstrate the potential of the bolometric technique for exotic double-$\beta$ decay searches. These promising results motivate the interest in investigating these BSM processes in the next-generation CUPID experiment. This study extensively analyzed the main limitations in sensitivity. In particular, the theoretical uncertainties on the \nbb decay spectral shape, the uncertainty on the presence of pure $\beta$-emitters in crystals, and the small statistics are the primary limiting factors. In the future, improvements in NME calculations can further constrain the $\xi$ parameters of the \nbb decay improved description. For the next-generation CUPID experiment, the possibility of measuring the $^{90}$Sr and \K concentration in \lmo crystals with a sensitivity $\lesssim 10^{-20}$ g/g will be extremely helpful in overcoming the problem of pure $\beta$-decays.

\section{Acknowledgments}
This work has been performed in the framework of the CUPID-1 (ANR-21-CE31-0014) and LUMINEU programs, funded by the Agence Nationale de la Recherche (ANR, France).
We acknowledge also the support of the P2IO LabEx (ANR-10-LABX0038) in the framework ''Investissements d'Avenir'' (ANR-11-IDEX-0003-01 – Project ''BSM-nu'') managed by ANR, France.
The help of the technical staff of the Laboratoire Souterrain de Modane and of the other participant laboratories is gratefully acknowledged. 
We thank the mechanical workshops of LAL (now IJCLab) for the detector holders fabrication and CEA/SPEC for their valuable contribution in the detector conception. 
V.V. Kobychev, O.G. Polischuk, and M.M. Zarytskyy were supported in part by the National Research Foundation of Ukraine Grant No. 2023.03/0213. O.G. Polischuk was supported in part by the project “Investigations of rare nuclear processes” of the program of the National Academy of Sciences of Ukraine “Laboratory of young scientists”. J. Kotila is supported by Academy of Finland (Grant Nos. 3314733, 320062, 345869). F. \v{S}imkovic acknowledges support from the Slovak Research and Development Agency under Contract No. APVV-22-0413 and by the Czech Science Foundation (GA\v{C}R), project No. 24-10180S.
Additionally the work is supported by the Istituto Nazionale di Fisica Nucleare (INFN) and by the EU Horizon2020 research and innovation program under the Marie Sklodowska-Curie Grant Agreement No. 754496. This work is also based on support by the US Department of Energy (DOE) Office of Science under Contract Nos. DE-AC02-05CH11231, and by the DOE Office of Science, Office of Nuclear Physics under Contract Nos. DE-FG02-08ER41551, DE-SC0011091; by the France-Berkeley Fund, the MISTI-France fund and  by the Chateau-briand Fellowship of the Office for Science \& Technology of the Embassy of France in the United States. This research used resources of the National Energy Research Scientific Computing Center (NERSC) and the IN2P3 Computing Centre.
This work makes use of the DIANA data analysis software and the background model based on JAGS,  developed by the Cuoricino, CUORE, LUCIFER, and CUPID-0 Collaborations. 
\noindent
Russian and Ukrainian scientists have given and give crucial contributions to CUPID-Mo. For this reason, the CUPID-Mo collaboration is particularly sensitive to the current situation in Ukraine. The position of the collaboration leadership on this matter, approved by majority, is expressed at \url{https://cupid-mo.mit.edu/collaboration#statement}. The majority of the work described here was completed before February 24, 2022.

\bibliographystyle{apsrev4-1}       
\bibliography{biblio}

\end{document}